\documentclass[runningheads]{cl2emult}

\usepackage{makeidx}  
\usepackage{graphicx} 
\usepackage{subeqnar} 
                      
\usepackage{multicol}
\usepackage{cropmark}
                      
\usepackage{eso}      
\makeindex

\def\Msun{\hbox{$\rm\thinspace M_{\odot}$}}

\begin{document}

\title*{Supermassive Black Holes: Their Formation,  \protect\newline and Their Prospects as Probes of Relativistic Gravity}

\toctitle{Supermassive Black Holes: Their Formation,  \protect\newline and Their Prospects as Probes of Relativistic Gravity}

\titlerunning{Supermassive Black Holes}

\author{Martin J. Rees\inst{}
}

\authorrunning{Martin J. Rees}

\institute{Institute of Astronomy\\Madingley Road, Cambridge, CB3 0HA, UK}

\maketitle            

\begin{abstract}
The existence of supermassive collapsed objects in the
cores of most galaxies poses still-unanswered questions. First, how
did they form, and how does their mass depend on the properties of the
host galaxy? Second, can observations probe the metric in the
strong-field domain, testing whether it indeed agrees with the Kerr
geometry predicted by general relativity (and, if so, what the spin
is)?
\end{abstract}

\section{Introduction}

Compact dark objects, with deep gravitational
    potential wells, seem to lurk in most galactic centres; but
    current evidence cannot `diagnose' the metric in the innermost
    region where Newtonian approximations break down.  Several other
    speakers have described the status of the observations, as well as
    some aspects of theoretical models.  This written text addresses
    two issues.  How did supermassive holes form? And do they have
    Schwarzschild/Kerr metrics, thereby offering real prospects of
    testing our theories of strong-field gravity.
 
     It has long been suspected that supermassive holes are implicated
in the power output from active galactic nuclei (AGNs), and in the
production of relativistic jets that energise strong radio
sources. But the demography of these massive holes has been clarified
by studies of relatively nearby galaxies: the centres of most of these
display either no activity or a rather low level, but most seem to
harbour dark central masses. Recent observational progress brings into
sharper focus the question of how and when supermassive black holes
formed, and how this process relates to galaxy formation.

   There are now two spectacularly convincing cases of massive
collapsed objects in nearby galaxies. The first, in the peculiar
spiral NGC 4258, has been revealed by amazingly precise mapping of gas
motions via the 1.3 cm maser-emission line of H$_2$O. [1,2].  The
spectral resolution of this microwave line is high enough to pin down
the velocities with accuracy of 1 km/sec.  The Very Long Baseline
Array achieves an angular resolution better than 0.5 milliarc seconds
(100 times sharper than the HST, as well as far finer spectral
resolution of velocities!). These observations have revealed, right in
NGC 4258's core, a disc with rotational speeds following an exact
Keplerian law around a compact dark mass.  The inner edge of the
observed disc is orbiting at 1080 km/sec. It would be impossible to
circumscribe, within its radius, a stable and long-lived star cluster
with the inferred mass of $3.6 \times 10^7 \Msun$.

   The second utterly convincing candidate lies in our own Galactic
Centre.  Most nearby large galaxies seem to harbour massive central
holes, so our own would seem underendowed if it did not have one
too. Some have advanced this view for many years (eg ref [3]). Also,
an unusual radio source has long been known to exist right at the
dynamical centre of our Galaxy, which can be interpreted in terms of
accretion onto a massive hole [4-6]. Direct evidence used to be
ambiguous because intervening gas and dust in the plane of the Milky
Way prevents us from getting a clear optical view of the central
stars, as we can in, for instance, M 31. A great deal was known about
gas motions, from radio and infrared measurements, but these were hard
to interpret because gas does not move ballistically like stars, but
can be influenced by pressure gradients, stellar winds, and other
non-gravitational influences.

    The situation was transformed by remarkable observations of stars
in the near infrared band, where obscuration by intervening material
is less of an obstacle. These are presented by Ekhart and by Ghez at
this meeting. The speeds scale as $r^{-1/2}$ with distance from the centre,
consistent with a hole of mass $2.5 \times 10^6 \Msun$.

        As other speakers will discuss, there is a crude
proportionality between the hole's mass and that of the central bulge
or spheroid in the stellar distribution (which is of course the
dominant part of an elliptical galaxy, but only a subsidiary component
of a disc system like M31 or our own Galaxy).  But how did the holes
form?

\section{AGN Demography and Black Hole Formation}
      
     Many of the faint smudges visible in the Hubble Deep Field [7]
are galaxies with redshifts of order 3, being viewed at (or even
before) the era when their spheroids formed.  Physical conditions in
the central potential wells of young and gas-rich galaxies should be
propitious for black hole formation, and such processes are presumably
connected with high-$z$ quasars.  It now seems clear that most galaxies
that existed at $z= 3$ would have participated subsequently in a series
of mergers; giant present-day elliptical galaxies are the outcome of
such mergers.  Any black holes already present would tend to spiral
inwards, and coalesce (unless a third body fell in before the merger
was complete, in which case a Newtonian slingshot could eject all
three: a binary in one direction; the third, via recoil, in the
opposite direction).
  
  The issues are  then:

{(a)} how much does a black hole grow (and how much electromagnetic
energy does it radiate) at each stage? and    
 
{(b)} how far up the `merger tree' did the first massive holes form? A
single big galaxy can be traced back to the stage when it was in
dozens of smaller components with individual internal velocity
dispersions as low as 20 km/sec.  Did central black holes form even in
these small and weakly bound systems? If so, they could have
coalesced, in a hierarchical fashion, during subsequent mergers.
 
     Perhaps black holes form with the same efficiency in small
galaxies (with shallow potential wells), or maybe their formation had
to await the buildup of substantial galaxies with deeper potential
wells (i.e. with $V$ above some threshold). This issue is important for
the detectability of high-$z$ miniquasars; it also determines whether
the ionizing UV background at high redshifts has a nonthermal
component that is able to ionize He as well as H.

    The actual formation mechanism is still uncertain. More than 20
years ago, I presented a `flow diagram' [8] which carried the message
that it seemed likely -- indeed almost inevitable - that large masses
would collapse in galactic centres: there was indeed a variety of
possible routes.  We have now got used to the idea that black holes
exist within most galaxies, but it is rather depressing that we still
cannot decide which formation route is most likely.

     One possibility is that the gas in a `proto-spheroid' does not
all break up into stellar-mass condensations, but that a supermassive
star forms, which then collapses.  As the gas evolved (through loss of
energy and angular momentum) to higher densities and more violent
internal dissipation, radiation pressure would prevent fragmentation,
and puff it up into a single superstar [9,10].  Ordinary star
formation may be suppressed even at less extreme densities -- i.e.
before the gas has become a single superstar -- by other effects. For
example, a magnetic field, even if not dynamically important overall,
could inhibit fragmentation, especially because the free-electron
concentration is unlikely to fall low enough to permit ambipolar
diffusion, whereby the magnetic flux can escape from protostars in
present-day molecular clouds.

       Once a large mass of gas started to behave like a single
superstar, it would continue to contract and deflate. Some mass would
inevitably be shed, carrying away angular momentum, but the remainder
would undergo complete gravitational collapse. This could be a
substantial fraction -- for example, if 10 percent of the mass had to
be shed in order to allow contraction by a factor of 2, about 20
percent could form a black hole [10,11].

          The mass of the hole would depend on that of its host
galaxy, though not necessarily via an exact proportionality: the
angular momentum of the protogalaxy and the depth of its central
potential well are relevant factors too.  Firmer and more quantitative
conclusions will have to await elaborate numerical simulations. But it
certainly seems in no way implausible that massive black holes form
directly from gas (some, albeit, already processed through stars),
perhaps after a transient phase as a supermassive object.

   However, we cannot exclude the alternative `scenario' where a
massive star builds up within a dense central cluster of ordinary
stars.  The most detailed calculations were done by Quinlan and
Shapiro ([12] and other references cited therein). These authors
showed that stellar coalescence, followed by the segregation of the
resultant high-mass stars towards the centre, could trigger runaway
evolution without (as earlier and cruder work had suggested) requiring
implausible initial starting points.  It would be well worthwhile
extending these simulations to a wider range of initial conditions,
and also to follow the build-up from stellar masses to supermassive
object.

      It is worth noting, incidentally, that whereas activity in low-$z$
galaxies may be correlated with some unusual disturbance due to a
tidal encounter or merger, this may not be the right way to envisage
the more common high-$z$ quasars, since almost all high-$z$ galaxies are
`disturbed', in the sense that they are nearly always experiencing a
merger or disturbance that is sufficient to perturb axisymmetry or to
trigger a large inflow of gas.

    The most massive black holes would have gained mass through a
succession of mergers, as well as through accretion of gas at each
stage.  Haehnelt and Kauffmann ([13], and these proceedings) have
modelled this in the context of semi-analytic schemes for galaxy
evolution, and have achieved a good fit with the luminosity function
and $z$-dependence of quasars.

\section{Do the Candidate Holes  Obey the Kerr Metric?}

\subsection{Probing  near the hole}

     The observed molecular disc in NGC 4258 lies a long way out: at
around $10^5$ gravitational radii.  We can exclude all conventional
alternatives (dense star clusters, etc); however, the measurements
tell us nothing about the central region where gravity is strong --
certainly not whether the putative hole actually has properties
consistent with the Kerr metric. The stars closest to our Galactic
Centre likewise lie so far out from the putative hole (their speeds
are less than 1 percent that of light) that their orbits are
essentially Newtonian.

       We can infer from AGNs that `gravitational pits' exist, which
must be deep enough to allow several percent of the rest mass of
infalling material to be radiated from a region compact enough to vary
on timescales as short as an hour. But we still lack quantitative
probes of the relativistic region. We believe in general relativity
primarily because it has been resoundingly vindicated in the weak
field limit (by high-precision observations in the Solar System, and
of the binary pulsar) -- not because we have evidence for black holes
with the precise Kerr metric.
   
        Relativists would seize eagerly on any relatively `clean'
probe of the strong-field domain.  The emission from most accretion
flows is concentrated towards the centre, where the potential well is
deepest and the motions fastest.  Such basic features of the
phenomenon as the overall efficiency, the minimum variability
timescale, and the possible extraction of energy from the hole itself
all depend on inherently relativistic features of the metric -- on
whether the hole is spinning or not, how it is aligned, etc.  But the
data here are imprecise and `messy'.  We would occasionally expect to
observe, even in quiescent nuclei, the tidal disruption of a
star. Exactly how this happens would depend on distinctive precession
effects around a Kerr metric, but the gas dynamics are so complex that
even when a flare is detected it will not serve as a useful diagnostic
of the metric in the strong-field domain. There are however several
encouraging new possibilities.

 \subsection{X-ray  spectroscopy of accretion flows}

       Optical spectroscopy tells us a great deal about the gas in
AGNs.  However, the optical spectrum originates quite far from the
hole. This is because the innermost regions would be so hot that their
thermal emission emerges as more energetic quanta.  X-rays are a far
more direct probe of the relativistic region.  The appearance of the
inner disc around a hole, taking doppler and gravitational shifts into
account, along with light bending, was first calculated by Bardeen and
Cunningham [14] and subsequently by several others (eg [15]). There is
of course no hope (until X-ray interferometry is developed) of
actually `imaging' these inner discs. However, the large
frequency-shifts could reveal themselves spectroscopically ---
substantial gravitational redshifts would be expected, as well as
large doppler shifts [15].  Until recently, the energy resolution and
sensitivity of X-ray detectors was inadequate to permit spectroscopy
of extragalactic objects. The ASCA X-ray satellite was the first with
the capability to measure emission line profiles in AGNs.  There is
already one convincing case [16] of a broad asymmetric emission line
indicative of a relativistic disc, and others should soon follow.  The
value of (a/m) can in principle be constrained too, because the
emission is concentrated closer in, and so displays larger shifts, if
the hole is rapidly rotating, and there is some evidence that this
must be the case in MCG -6-30-15 [17].

   The Chandra and XMM X-ray satellites should be able to extend and
refine these studies; they may offer enough sensitivity, in combination
with time-resolution, to study flares, and even to follow a `hot spot'
on a plunging orbit.

   The swing in the polarization vector of photon trajectories near a
hole was long ago suggested [18] as another diagnostic; but this is
still not feasible because X-ray polarimeters are far from capable of
detecting the few percent polarization expected.

\subsection{The Blandford-Znajek process}

    Blandford and Znajek [19] showed that a magnetic field threading a
hole (maintained by external currents in, for instance, a torus) could
extract spin energy, converting it into directed Poynting flux and
electron-positron pairs.  Can we point to objects where this is
definitively happening?  The giant radio lobes from radio galaxies
sometimes spread across millions of lightyears -- $10^{10}$ times larger
than the hole itself. If the Blandford-Znajek process is really going
on, these huge structures may be the most direct manifestation of an
inherently relativistic effect around a Kerr hole.

    Jets in some AGNs definitely have Lorentz factors exceeding
10. Moreover, some are probably Poynting-dominated, and contain pair
(rather than electron-ion) plasma. But there is still no compelling
reason to believe that these jets are energised by the hole itself,
rather than by winds and magnetic flux `spun off' the surrounding
torus. The case for the Blandford-Znajek mechanism would be
strengthened if baryon-free jets were found with still higher Lorentz
factors, or if the spin of the holes could be independently measured,
and the properties of jets turned out to depend on (a/m).

  The process cannot dominate unless either the field threading the
hole is comparable with that in the orbiting material, or else the
surrounding material radiates with low radiative efficiency. These
requirements cannot be ruled out, though there has been recent
controversy about how plausible they are. (It may be worth noting that
the Blandford-Znajek effect could also be important in the still more
extreme context of gamma-ray bursts, where a newly formed hole of a
few solar masses could be threaded by a field exceeding $10^{15}$  G.)

\subsection{What is the expected spin?}

  The spin of a hole affects the efficiency of `classical' accretion
processes; the value of a/m also determines how much energy is in
principle extractable by the Blandford-Znajek effect.  Moreover, the
orientation of the spin axis may be important in relation to jet
production, etc.

    Spin-up is a natural consequence of prolonged disc-mode accretion:
any hole that has (for instance) doubled its mass by capturing
material that is all spinning the same way would end up with a/m being
at least 0.5.  A hole that is the outcome of a merger between two of
comparable mass would also, generically, have a substantial spin. On
the other hand, if it had gained its mass from capturing many low-mass
objects (holes, or even stars) in randomly-oriented orbits, a/m would
be small.

\subsection{Precession and alignment}

   Most of the literature on
gas dynamics around Kerr holes assumes that the
flow is axisymmetric. This assumption is motivated not just by
simplicity, but by the expectation that Lense-Thirring precession
would impose axisymmetry close in, even if the flow further out were
oblique and/or on eccentric orbits.  Plausible-seeming arguments,
dating back to the pioneering 1975 paper by Bardeen and Petterson [20],
suggested that the alignment would occur, and would extend out to a
larger radius if the viscosity were low because there would be more
time for Lense-Thirring precession to act on inward-spiralling gas.
However, later studies, especially by Pringle, Ogilvie, and their
associates, have shown that naive intuitions can go badly awry. The
behaviour of the `tilt' is much more subtle; the effective viscosity
perpendicular to the disc plane can be much larger than in the
plane. In a thin disc, the alignment effect is actually weaker when
viscosity is low. What happens in a thick torus is a still unclear,
and will have to await 3-D gas-dynamical simulations.

        The orientation of a hole's spin and the innermost flow
patterns could have implications for jet alignment. An important paper
by Pringle and Natarajan [21] shows that `forced precession' effects
due to torques on a disc can lead to swings in the rotation axis that
are surprisingly fast (i.e. on timescales very much shorter than the
timescale for changes in the hole's mass).

\subsection{Stars in relativistic orbits?}

    Gas-dynamical phenomena are complicated because of viscosity,
magnetic fields etc. It would be nice to have a `cleaner' and more
quantitative probe of the strong-field regime: for instance, a small
star orbiting close to a supermassive hole.  Such a star would behave
like a test particle, and its precession would probe the metric in the
`strong field' domain.  These interesting relativistic effects, have
been computed in detail by Karas and Vokrouhlicky [22,23]. Would we
expect to find a star in such an orbit?

   An ordinary star certainly cannot get there by the kind of `tidal
capture' process that can create close binary star systems. This is
because the binding energy of the final orbit (a circular orbit with
the same angular momentum as an initially near-parabolic orbit with
pericentre at the tidal-disruption radius) would have to be dissipated
within the star, and that cannot happen without destroying it.  Syer,
Clarke and Rees [24] pointed out, however, that an orbit can be
`ground down' by successive impacts on a disc (or any other resisting
medium) without being destroyed: the orbital energy then goes almost
entirely into the material knocked out of the disc, rather than into
the star itself. Other constraints on the survival of stars in the
hostile environment around massive black holes -- tidal dissipation
when the orbit is eccentric, irradiation by ambient radiation, etc --
are explored by Podsiadlowski and Rees [25], and King and Done [26].
They can be thought of as close binary star systems with extreme mass
ratios.

   These stars would not be directly observable, except maybe in our
own Galactic Centre.  But they might have indirect effects: such a
rapidly-orbiting star in an active galactic nucleus could signal its
presence by quasiperiodically modulating the AGN emission.

\subsection{Gravitational-wave capture of compact stars} 

   Neutron stars or white dwarfs circling close to supermassive black
holes would be impervious to tidal dissipation, and would have such a
small geometrical cross section that the `grinding down' process would
be ineffective too. On the other hand, because they are small they can
get into very tight orbits by straightforward stellar-dynamical
processes. For ordinary stars, the `point mass' approximation breaks
down for encounter speeds above 1000 km/s -- physical collisions are
then more probable than large-angle deflections. But there is no
reason why a `cusp' of tightly bound compact stars should not extend
much closer to the hole.  Neutron stars or white dwarfs could exchange
orbital energy by close encounters with each other until some got
close enough that they either fell directly into the hole, or until
gravitational radiation became the dominant energy loss. When stars
get very close in, gravitational radiation losses become significant, and
tend to circularise an elliptical orbit with small pericentre. Most
such stars would be swallowed by the hole before circularisation, because the
angular momentum of a highly eccentric orbit `diffuses' faster than
the energy does due to encounters with other stars, but some would get
into close circular orbits [27,28].

    A compact star is less likely than an ordinary star in similar
orbit to `modulate' the observed radiation in a detectable way.  But
the gravitational radiation (almost periodic because the dissipation
timescale involves a factor $(M_{\rm hole}/m^*)$) would be detectable.

\subsection{Scaling laws and `microquasars'}

     Two galactic X-ray sources that are believed to involve black
holes generate double radio structures that resemble miniature
versions of the classical extragalactic strong radio sources. These
are discussed in the paper by Mirabel. The jets have been found to
display apparent superluminal motions across the sky, indicating that,
like the extragalactic radio sources, they contain plasma that is
moving relativistically.

  There is no reason to be surprised by this analogy between phenomena
on very different scales. Indeed, the physics is exactly the same,
apart from very simple scaling laws. 
If we define $l = L/L_{\rm Ed}$ and $\dot m = \dot M/
\dot M_{\rm crit}$, where $\dot M_{\rm crit} = L_{\rm Ed}/c^2$, then for a given value of $\dot m$, the flow
pattern may be essentially independent of $M$. Linear scales and
timescales, of course, are proportional to M, and densities in the
flow scale as $M^{-1}$.  The physics that amplifies and tangles any
magnetic field may be scale-independent, and the field strength B
scales as $M^{-1/2}$. So the bremsstrahlung or synchrotron cooling
timescales go as $M$, implying that $t_{\rm cool}/t_{\rm dyn}$ is insensitive to $M$ for a given $\dot m$. So
also are ratios involving, for instance, coupling of electron and ions
in thermal plasma. Therefore, the efficiencies and the value of $l$ are
insensitive to $M$, and depend only on $\dot m$.  Moreover, the form of the
spectrum, for given $\dot m$,  depends on M only rather insensitively (and in a manner that
is easily calculated).

  The kinds of accretion flow inferred in, for instance, M87, giving
rise to a compact radio and X-ray source, along with a relativistic
jet, could operate just as well if the hole mass was lower by a
hundred million, as in the galactic LMXB sources.  So we can actually
study the same processes involved in AGNs in microquasars close at
hand within our own galaxy. And these miniature sources may allow us
to observe a simulacrum of the entire evolution of a strong
extragalactic radio source, its life-cycle speeded up by a similar factor.

\subsection{Discoseismology}

   Discs or tori that are maintained by steady flow into a black hole
can support vibrational modes [29-31].  The frequencies of these modes
can, as in stars, serve as a probe for the structure of the inner disc
or torus. The amplitude depends on the importance of pressure, and
hence on disc thickness; how they are excited, and the amplitude they
may reach, depends, as in the Sun, on interaction with convective
cells and other macroscopic motions superimposed on the mean flow. But
the frequencies of the modes can be calculated more reliably. In
particular, the lowest g-mode frequency is close to the maximum value
of the radial epicyclic frequency $k$.  This epicyclic frequency is, in
the Newtonian domain, equal to the orbital frequency. It drops to zero
at the innermost stable orbit. It has a maximum at about $9GM/c^2$ for a
Schwarzschild hole; for a Kerr hole, $k$ peaks at a smaller radius (and
a higher frequency for a given $M$). The frequency is 3.5 times higher
for $(a/m)=1$ than for the Schwarzschild case.

   Novak and Wagoner [31] pointed out that these modes may cause an
observable modulation in the X-ray emission from galactic black hole
candidates.  Just such effects have been seen in GRS 1915+105
[32]. The amplitude is a few percent (somewhat larger at harder X-ray
energies) suggesting that the oscillations involve primarily the
hotter inner part of the disc. The fluctuation spectrum shows a peak
in Fourier space at around 67 Hz. This frequency does not change even
when the X-ray luminosity doubles, suggesting that it relates to a
particular radius in the disc. If this is indeed the lowest g-mode,
and if the simple disc models are relevant, then the implied mass is
$10.2 \Msun$ for Schwarzschild, and $35 \Msun$ for a `maximal Kerr' hole (Nowak et
al 1997). The mass of this system is not well known. However, this
technique offers the exciting prospect of inferring (a/m) for holes
whose masses are independently known.

    GRS 1915+105 is one of the objects with superluminal radio
jets. The simple scaling arguments section 3.8. imply that the AGNs which
it resembles might equally well display oscillations with the same
cause. However, the periods there  would be measured in days, rather than
fractions of a second.

\section{Gravitational Radiation as a Probe}

\subsection{Gravitational waves from newly-forming massive holes?}

     The gravitational radiation from black holes offers impressive
tests of general relativity, involving no physics other than that of
spacetime itself.

     At first sight, the formation of a massive hole from a monolithic
collapse might seem an obvious source of strong wave pulses.  The
wave emission would be maximally intense and efficient if the holes
formed on a timescale as short as $(r_g/c)$, where $r_g= (GM/c^2)$ --
something that might happen if they built up via coalescence of
smaller holes (cf ref [12]).

       If, on the other hand, supermassive black holes formed as
suggested in section 2 -- directly from gas (some, albeit, already
processed through stars), perhaps after a transient phase as a
supermassive object -- then the process may be too gradual to yield
efficient gravitational radiation.  That is because post-Newtonian
instability is triggered at a radius $r_i \gg r_g$.  Supermassive stars
are fragile because of the dominance of radiation pressure: this
renders the adiabatic index only slightly above 4/3 (by an amount of
order $10(M/\Msun)^{-1/2})$. Since = 4/3 yields neutral stability in Newtonian
theory, even the small post-Newtonian corrections then destabilize
such `superstars'.  The characteristic collapse timescale when
instability ensues is longer than $r_g/c$ by the 3/2 power of the
collapse factor.

    The post-Newtonian instability is suppressed by rotation.  A
differentially rotating supermassive star could in principle support
itself against post-Newtonian instability until it became very tightly
bound.  It could then perhaps develop a a bar-mode instability and
collapse within a few dynamical times. To achieve this tightly-bound
state without drastic mass loss, the object would need to have deflated over a long timescale,
losing energy at no more than the Eddington rate.
 
       The formation of a hole `in one go' from a supermassive star is
an unpromising source of gravitational waves. On the other hand,
strong signals are expected when already-formed holes coalesce, as the
aftermath of mergers of their host galaxies.

       The gravitational waves associated with supermassive holes
would be concentrated in a frequency range around a millihertz -- too
low to be accessible to ground-based detectors, which lose sensitivity
below 100 Hz , owing to seismic and other background
noise. Space-based detectors are needed.  One such, proposed by the
European Space Agency, is the Laser Interferometric Spacecraft (LISA)
-- three  spacecraft on solar orbit, configured as a triangle, with 
sides  of 5 million km long whose length is monitored by laser
interferometry.
    
\subsection{Gravitational waves from coalescing supermassive holes.}

     The guaranteed sources of really intense gravitational waves in
LISA's frequency range would be coalescing supermassive black
holes. Many galaxies have experienced a merger since the epoch $z > 2$
when, according to `quasar demography' arguments  they
acquired central holes.  The holes in the two merging galaxies would
spiral together, emitting, in their final coalescence, up to $\sim 10$  per
cent of their rest mass as a burst of gravitational radiation in a
timescale of only a few times $r_g/c$. These pulses would be so strong
that LISA could detect them with high signal-to-noise even from large
redshifts.  Whether such events happen often enough to be interesting
can to some extent be inferred from observations (we see many galaxies
in the process of coalescing), and from simulations of the
hierarchical clustering process whereby galaxies and other cosmic
structures form.  Haehnelt [33 and later references] has calculated
the merger rate of the large galaxies believed to harbour supermassive
holes: it is only about one event per century, even out to redshifts $z
= 4$.  However, big galaxies are probably the outcome of many
successive mergers. As discussed in Section 2, we still have no direct
evidence -- nor firm theoretical clues -- on whether these small
galaxies harbour black holes (nor, if they do, of what the hole masses
typically are). However it is certainly possible that enough holes of
(say) $10^5 \Msun$ lurk in small early-forming galaxies to yield, via
subsequent mergers, more than one event per year detectable by LISA.

      LISA is potentially so sensitive that it could detect the
nearly-periodic waves from stellar-mass objects orbiting a 
$10^5 -
10^6 \Msun$ hole, even at a range of a hundred Mpc, despite the $m^*/M_{\rm hole}$ factor
whereby the amplitude is reduced compared with the coalescence of two
objects of comparable mass $M_{\rm hole}$. The stars in the observed `cusps' around
massive central holes in nearby galaxies are of course (unless almost
exactly radial) on orbits that are far too large to display
relativistic effects.  Occasional captures into relativistic orbits
can come about by dissipative processes -- for instance, interaction
with a massive disc [24,34]. But unless the hole mass were above $10^8
\Msun$ (in which case the waves would be at too low a frequency for LISA
to detect), solar-type stars would be tidally disrupted before getting
into relativistic orbits. Interest therefore focuses on compact stars,
for which dissipation due to tidal effects or drag is less
effective. As described in Section 3.7, compact stars may get captured
as a result of gravitational radiation, which can gradually `grind
down' an eccentric orbit with close pericenter passage into a
nearly-circular relativistic orbit. The long quasi-periodic wave
trains from such objects, modulated by orbital precession (cf refs
[22,23]) in principle carries detailed information about the metric.
 
     The attraction of LISA as an `observatory' is that even
conservative assumptions lead to the prediction that a variety of
phenomena will be detected.  If there were many massive holes not
associated with galactic centres (not to mention other speculative
options such as cosmic strings), the event rate would be much
enhanced. Even without factoring on an `optimism factor' we can be
confident that LISA will harvest a rich stream of data.

     LISA is now being actively studied both in Europe and the US. If
funded jointly by ESA and NASA, it could fly within ten years.

\subsection{Gravitational-wave recoil}

     Is there any way of learning, before that date, something about
gravitational radiation?  The dynamics (and gravitational radiation)
when two holes merge has so far been computed only for cases of
special symmetry. The more general problem -- coalescence of two Kerr
holes with general orientations of their spin axes relative to the
orbital angular momentum -- is a `grand challenge' computational
project being tackled at the MPI in Potsdam, and at US centres. When
this challenge has been met (and it will almost certainly not take all
the time until LISA flies) we shall find out not only the
characteristic wave form of the radiation, but the recoil that arises
because there is a net emission of linear momentum.

   There would be a recoil due to the non-zero net linear momentum
carried away by gravitational waves in the coalescence. If the holes
have unequal masses, a preferred longitude in the orbital plane is
determined by the orbital phase at which the final plunge occurs.  For
spinning holes there may, additionally, be a rocket effect
perpendicular to the orbital plane, since the spins break the mirror
symmetry with respect to the orbital plane. [35]

   The recoil is a strong-field gravitational effect which depends
essentially on the lack of symmetry in the system.  It can therefore
only be properly calculated when fully 3-dimensional general
relativistic calculations are feasible.  The velocities arising from
these processes would be astrophysically interesting if they were
enough to dislodge the resultant hole from the centre of the merged
galaxy, or even eject it into intergalactic space. This recoil could
displace the hole from the centre of the merged galaxy -- it might
therefore be relevant to the low$-z$ quasars that seem to be
asymmetrically located in their hosts (and which may have been
activated by a recent merger).  Even galaxies that do not harbour a
central hole may, therefore, once have done so in the past. The core
of a galaxy that has experienced such an ejection event may retain
some trace of it (perhaps, for instance, an unusual profile), because
the energy transferred to stars via dynamical friction during the
merger process (cf [36]).
 
   The recoil might even be so violent that the merged hole breaks
loose from its galaxy and goes hurtling through intergalactic space.
This disconcerting thought should at least impress us with the reality
and `concreteness' of the extraordinary entities to whose discovery
Riccardo Giacconi contributed so much.

\section{Acknowledgements}

     I am grateful to several colleagues, especially Mitch Begelman,
Andy Fabian and Martin Haehnelt for discussions and collaboration on
topics mentioned here. I thank the Royal Society for support, and the
organisers of this conference for the opportunity to participate in
celebrating Riccardo Giacconi's extraordinary record of research and
scientific leadership.

\vfill

\begin{thebibliography}{7}

\addcontentsline{toc}{section}{References}

\bibitem{}Watson, W.D. and Wallin, B.K. 1994 Astrophys. J. (Lett) 432, L35
\bibitem{} Miyoshi, K. et al 1995 Nature 373, 127. 
\bibitem{} Lynden-Bell, D. and Rees, M.J. 1971 MNRAS 152, 461. 
\bibitem{} Rees, M.J. 1982 in `The Galactic Center' ed  G. Riegler and R.D. Blandford (A.I.P) p166.
\bibitem{} Melia, F. 1994 Astrophys. J. 426, 577
\bibitem{} Narayan, R., Yi, I, and Mahadevan, R. 1995 Nature 374, 623.
\bibitem{} Williams, R. et al 1996 Astron. J. 112, 1335.
\bibitem{} Rees, M.J. 1978, Observatory 98, 210.
\bibitem{} Rees, M.J. 1993 Proc. Nat. Acad. Sci 90, 4840
\bibitem{} Haehnelt M. and Rees, M.J. 1993 MNRAS 263, 168
\bibitem{} Baumgarte, T.W. and Shapiro, S.L. 1999 Ap. J. 526, 941.
\bibitem{} Quinlan, G.D. and Shapiro, S.L. 1990 Astrophys. J. 356, 483.
\bibitem{} Haehnelt, M. and Kauffmann G. 1999 MNRAS in press
\bibitem{} Bardeen, J. and Cunningham, J., 1972 Ap. J. 173, L137.
\bibitem{} N.E. White et al 1989 MNRAS 238, 729.
\bibitem{} Tanaka, Y. et al 1995 Nature 375, 659
\bibitem{} Iwasawa, I. et al 1999,  MNRAS  306, L191.
\bibitem{} Connors, P.A., Piran, T. and  and Stark, R.F.  1980 Ap. J 235, 224
\bibitem{} Blandford, R.D. and Znajek, R.L. 1977 MNRAS 179, 433
\bibitem{} Bardeen, J. and Petterson, J.A. 1975 Ap. J. 195, L65.
\bibitem{} Natarajan, P. and Pringle, J.E. 1999 Ap. J. in press.
\bibitem{} Karas, V., and Vokrouhlicky, D., 1993 MNRAS 265, 365
\bibitem{} Karas, V., and Vokrouhlicky, D., 1994 Astrophys. J 422, 208
\bibitem{} Syer, D., Clarke, C.J. and Rees, M.J. 1991 MNRAS 250, 505.
\bibitem{}  Podsiadlowski, P. and Rees, M.J. 1994 in `Evolution of X-ray binaries' 
ed S Holt and C.Day (AIP) p403.
\bibitem{} King, A.R. and Done. C., 1993 MNRAS 264, 388
\bibitem{} Hils, D. and Bender, P.L.  1995 Astrophys. J. (Lett) 445, L7 
\bibitem{} Sigurdsson, S. and Rees M.J. 1997 MNRAS 284, 318.
\bibitem{} Kato, S.   and Fukui, J.  1980 PASJ 32, 377 
\bibitem{} Novak, M.A.  and Wagoner, R.V. 1992 Astrophys. J. 393, 697
\bibitem{} Novak, M.A.  and Wagoner, R.V. 1993 Astrophys J. 418, 187
\bibitem{} Morgan, E., Remillard, R. and Greiner, J. 1996 IAU Circular No. 6392. 
\bibitem{} Haehnelt M. 1994 MNRAS 269,199
\bibitem{} Canizzo, J.K., Lee, H.M. and Goodman, J. 1990 Astrophys. J. 351, 38
\bibitem{} Redmount, I. and Rees, M.J. 1989 Comm. Astrophys. Sp. Phys 14, 185.
\bibitem{}  Ebisuzaki, T., Makino, J., and Okumura, S.K. 1991 Nature 354, 212
\end{thebibliography}
\end{document}